# Phases of planar QCD on the torus

**Rajamani Narayanan**

*Florida International University, Department of Physics, Miami, FL 33199*

*E-mail:* `rajamani.narayanan@fiu.edu`

**Herbert Neuberger***

*Rutgers University, Department of Physics and Astronomy, Piscataway, NJ 08855*

*E-mail:* `neuberg@physics.rutgers.edu`

At infinite $N$, continuum Euclidean $SU(N)$ gauge theory defined on a symmetrical four torus has a rich phase structure with phases where the finite volume system behaves as if it had infinite extent in some or all of the directions. In addition, fermions are automatically quenched, so planar QCD should be cheaper to solve numerically that full QCD. Large $N$ is a relatively unexplored and worthwhile direction of research in lattice field theory.

*Speaker.

## 1. Introduction

Sometimes plenary talks are given at a time well past the main conceptual progress on the topic. Not so this time: we have new results, but we feel that there are major new discoveries left to be made. The talk will reflect this by devoting a larger fraction of time to work in progress, things one cannot yet find on the archive. More than the news, we wish to convey that large $N$ is a new exciting research direction for lattice field theory.

Our collaborators on various projects are: J. Kiskis, A. González-Arroyo, L. Del Debbio, E. Vicari. Other interested parties are invited to join us, as the number of things to do by far exceeds what we can handle.

## 2. Why work on large $N$ ?

- Lattice large $N$ work will contribute to the search for a string representation of QCD. There has been progress at large $N$ in the context of supersymmetric models [1], the best known example being the AdS/CFT correspondence that applies to $\mathcal{N} = 4$ supersymmetric $SU(N)$ YM theory which is conformal and has no coupling constant renormalization. Since QCD is also approximately conformal, models are being built, incorporating some mechanism of conformal symmetry breaking in an approximate gravitational dual to QCD. These models are usually compared to real life QCD, but it would be more appropriate to compare them to planar QCD, since the curvature of the background needs to be small. By building a data base of facts about planar QCD, lattice field theory can contribute to this activity, and perhaps even point us in the direction of an exact dual description of planar QCD.

- There is a shortcut to $N = \infty$, known as "reduction" [2] which says that one can forget about going to the usual thermodynamic limit, so long as $N$ is large enough. In practice, one balances the system size against the magnitude of $N$ to attain maximal efficiency in approaching the planar limit.

- Even for massless quarks quenching is believed to be a valid approximation in the planar limit. As is well known, at finite $N$ the quenched massless theory is divergent [3], even calling into question some work done on the renormalization of fermionic observables in the context of quenched $SU(3)$: what is the meaning of the renormalized vacuum expectation value of the scalar fermion-bilinear at zero quark mass ? From large $N$ we know that such effects are of subleading order in $N$, and this is made evident by the standard parameterization of the quenched divergences. To be sure, at infinite $N$ the order of limits has to be $\lim_{m_q \to 0} \lim_{N \to \infty}$, otherwise the quenching divergences will not be handled correctly. This does not necessarily mean that every quantity needs to be explicitly calculated at several nonzero quark masses and subsequently extrapolated; in some cases there are indirect methods to obtain the desired quantity directly, but the definition in terms of a zero quark mass limit in the order described always holds.

- Large $N$ is a feasible problem for PC clusters of today, and can be useful both phenomenologically and as a case study. There are many quantities that have small $\frac{1}{N}$ corrections, and we need to learn which do not [4]. Clearly, when the dominating physics is that of diquark formation, we expect $\frac{1}{N}$ corrections to be large. Also, using large $N$ methods to study baryons seems a difficult proposition; for the time being, the focus is on mesons and there is plenty of $N = \infty$ physics there that is potentially relevant to QCD. Planar QCD is a good problem for PC clusters because in a parallel simulation the amount of computation per node increases as $N^3$, on account of matrix multiplications, while communication demands only increase as $N^2$, since the most complex objects have only two color indices. In practice the lattices are small, so only small clusters are useful for massively parallel simulations. In this respect large $N$ QCD is the ideal topic of research for the lattice field theorist that has access to a small cluster resource, operating in a small collaboration.

## 3. $N = \infty \approx N = 3$

Michael Teper and associates [5] have shown that the quantities

$$\frac{T_c}{\sqrt{\sigma}}, \frac{m_g^{xx}}{\sqrt{\sigma}}, \frac{P(T)}{P_{\text{free}}(T)}, \frac{\chi_{\text{top}}}{\sigma^2}$$

extrapolate smoothly to the planar limit, with corrections of order ten percent at $N = 3$. Here $\sigma$ is the string tension, $T_c$ is the finite temperature deconfinement transition, $m_g^{xx}$ is a shorthand notation for various glueball states, $P(T)$ is the pressure of the pure gauge system at temperature $T > T_c$ (while $P_{\text{free}}(T)$ is the same for a free gas a gluons) and $\chi_{\text{top}}$ is the topological susceptibility.

L. Del Debbio et.al. [6] have investigated the $\theta$-parameter dependence at large $N$ and see indications that it goes as proposed by Witten, namely that the vacuum energy as a function $\theta$ is given by periodically repeated truncated parabolas.

The above work has been going on for about seven years and has produced a wealth of results; it ought to be reviewed in a future plenary talk, because one could not do justice to it in the time allotted to this talk.

## 4. Plan and Summary

- Planar lattice QCD on an $L^4$ torus has 6 phases: 0h and [0-4]c [7]. Five of these phases, namely [0-4]c, survive in the continuum. In each phase one has a certain amount of large $N$ reduction. The 0h and 0c phases are totally insensitive to $L$.

- Planar QCD breaks chiral symmetry in 0c spontaneously and RMT works very well [8]:

$$\frac{1}{N}\frac{\langle\bar{\psi}\psi\rangle}{\sigma^{3/2}}|_{N=\infty} \approx \frac{1}{3}\frac{\langle\bar{\psi}\psi\rangle}{\sigma^{3/2}}|_{N=3}$$

- Large N reduction extends to mesons in 0c and the pion can be separated from the higher stable resonances [9]: We find that the naive formula

$$m_\pi^2 = am_q + bm_q^2$$

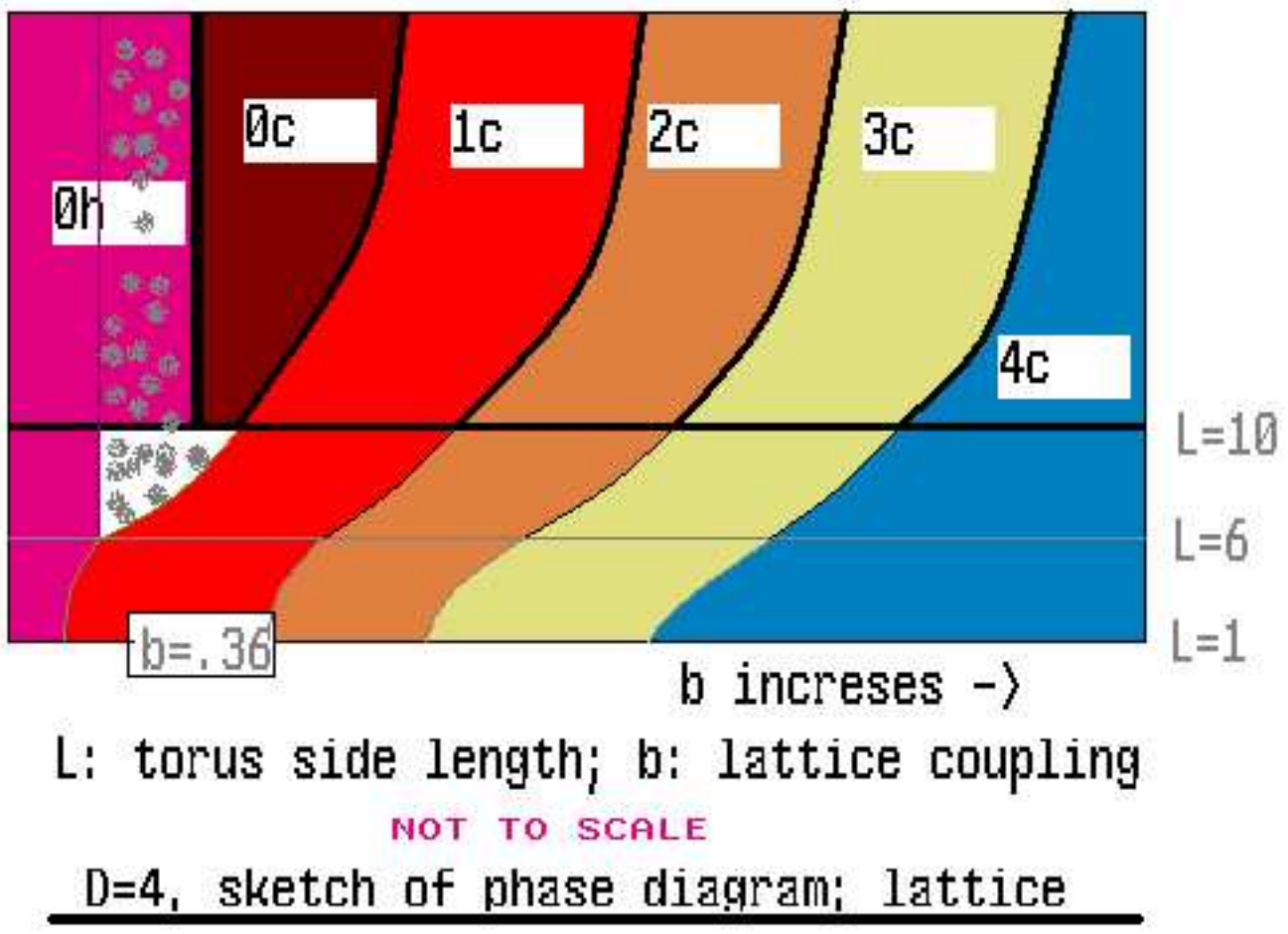

**Figure 1:** A schematic drawing of the phase diagram on the lattice

works over a large range of quark masses. Chiral logarithms are suppressed at large $N$. We also find a quantity that has large $\frac{1}{N}$ correction:

$$\frac{1}{N}\frac{f_\pi^2}{\sigma}\Big|_{N=\infty} \approx 2\,\frac{1}{3}\frac{f_\pi^2}{\sigma}\Big|_{N=3}$$

This correction, had it been known twenty years ago, would have changed the numbers in quite a few technicolor papers, but would have made little difference to the viability of technicolor ultimately.

- Chiral symmetry is restored in 1c and the Dirac operator develops a temperature dependent spectral gap.
- Twisting tricks hold the promise to reduce computation time significantly.
- This talk will end with some speculations.

## 5. Phase Structure on a finite Torus: Lattice and Continuum at $N = \infty$ [7]

Figure 1 shows the phase structure on a four dimensional toroidal lattice in a schematic manner. To date, only the 0c→1c transition has been mapped out. It is important to complete the mapping of the entire diagram. In the continuum, the 0h phase disappears. The physically interesting phase 0c extends by metastability into 0h, and this is used in practice to make simulations in 0$c$ at smaller $L$'s.

It is conjectured that all boundaries separating adjacent Xc phases will obey an asymptotic freedom dependence of the critical size on the coupling constant (that is, the usual exponential dependence dictated by the one loop beta function). The 0h→0c line is parallel to the y-axis and will not be intersected by a $b$=constant line at some large $b$. This implies that the 0$h$ phase, the single phase in which the large $N$ version of the strong coupling lattice expansion is useful, is separated by a first order phase transition from continuum physics. Although the 0h phase in itself depends on the specific form of the Wilson pure gauge action, it is suspected that no local lattice action which has a strong coupling phase will have this phase continuously connected to a continuum like phase in the planar limit. Thus, at infinite $N$, the strong coupling expansion is less of a credible argument for confinement than we are used to from our experience with $SU(2)$ and $SU(3)$ gauge theories.

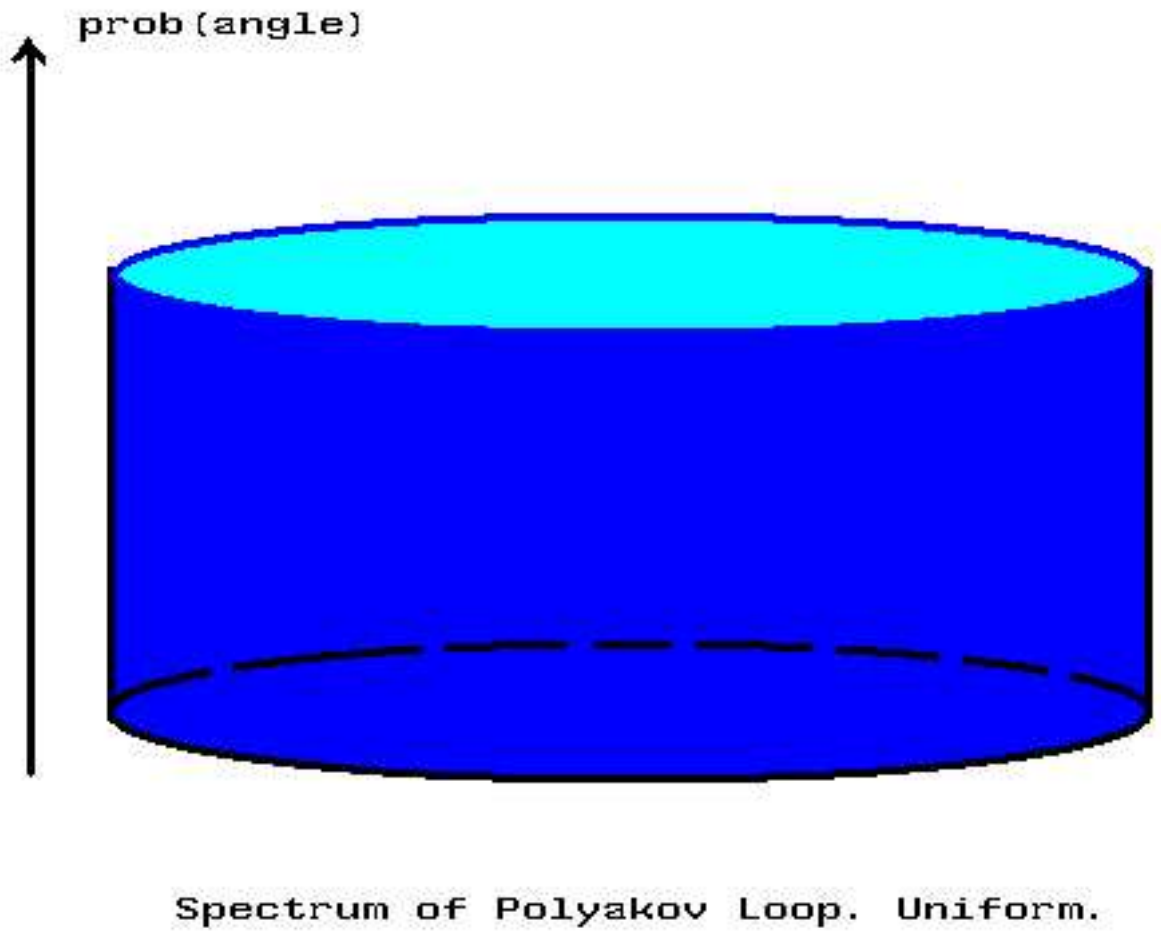

**Figure 2:** The spectrum of a Polyakov loop with no gap, obeying the $Z(N)$ symmetry in the respective direction

At small values of $L$, in particular $L = 1$, with a Wilson action, there is no 0c phase. In the past the entire focus was on the case $L = 1$ and the situation at higher $L$ was missed.

In three dimensions the situation is similar, only there is one less phase because the group that breaks sequentially as the torus is shrunk is $Z^3(N)$ rather than $Z^4(N)$.

The transitions between any two adjacent phases of type Xc occurs by the eigenvalue spectrum of the open Polyakov loop operator in some direction opening a gap. This is shown schematically in Figures 2 and 3.

In 0c all Polyakov Loops are uniform. In Xc some open gaps. In Figure 3 this is shown schematically. In a simulation one needs to take into account that ultraviolet fluctuations generate a renormalization of the trace of the Polyakov loop that goes like the exponent of its perimeter times the cutoff, in our case the lattice spacing. There are no corner divergences for Polyakov loops, since the loops are smooth. The perimeter divergence causes the trace to become very small numerically, and this is achieved by the eigenvalues of any unitary Polyakov loop operator having a spectrum which covers the entire unit circle almost uniformly. To see a gap opening one needs to eliminate the perimeter divergence. This is done by APE smearing the links and calculating the Polyakov loop operator from the smeared links. The exact amount of smearing is immaterial because the place where the $Z(N)$ breaks spontaneously is independent of the observable one uses. In Figure 4 actual data for such smeared and unsmeared Polyakov loops is shown. In some directions the $Z(N)$ symmetry is preserved and in others it is spontaneously broken, but this cannot be detected from the unsmeared loops, while it is evident from the smeared ones.

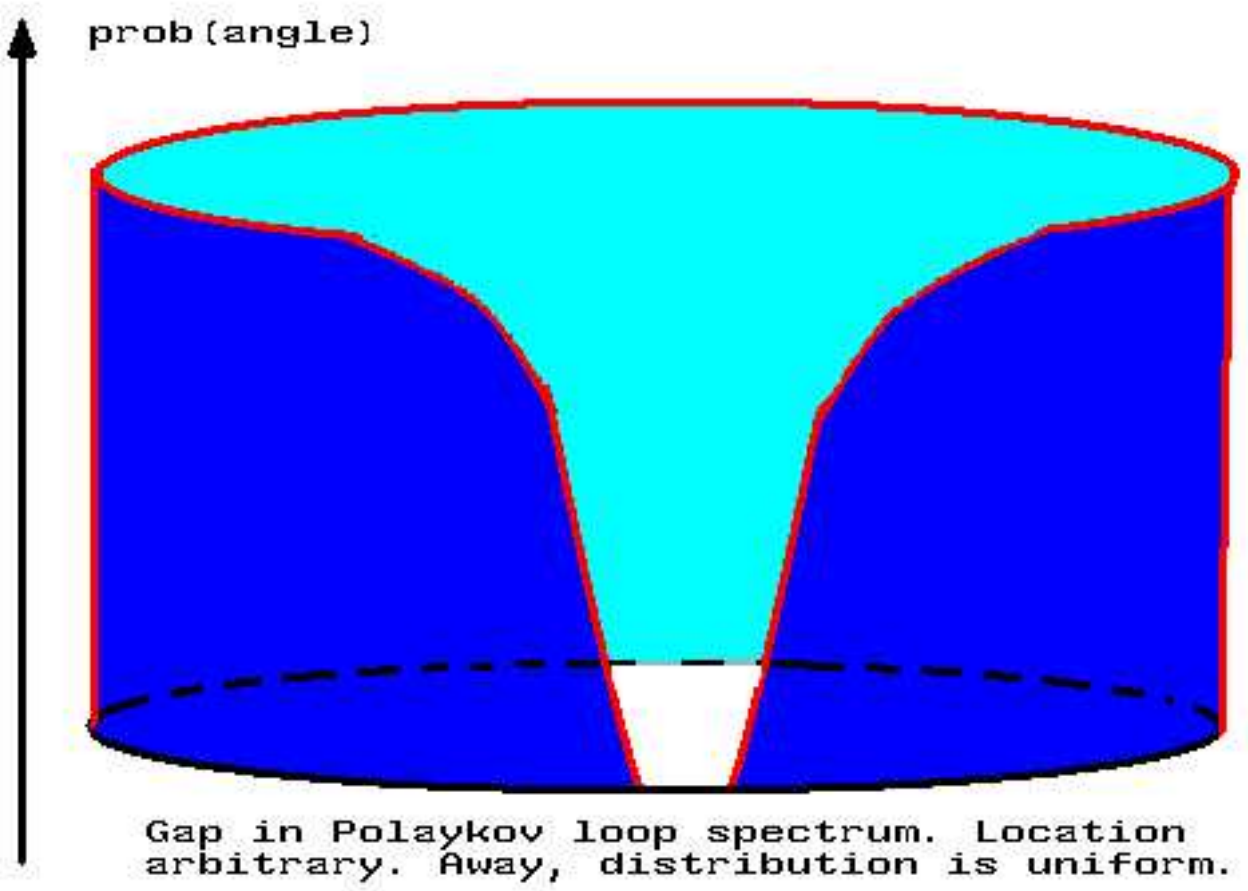

**Figure 3:** The spectrum of a Polyakov loop with a gap, spontaneously breaking the $Z(N)$ symmetry in the respective direction

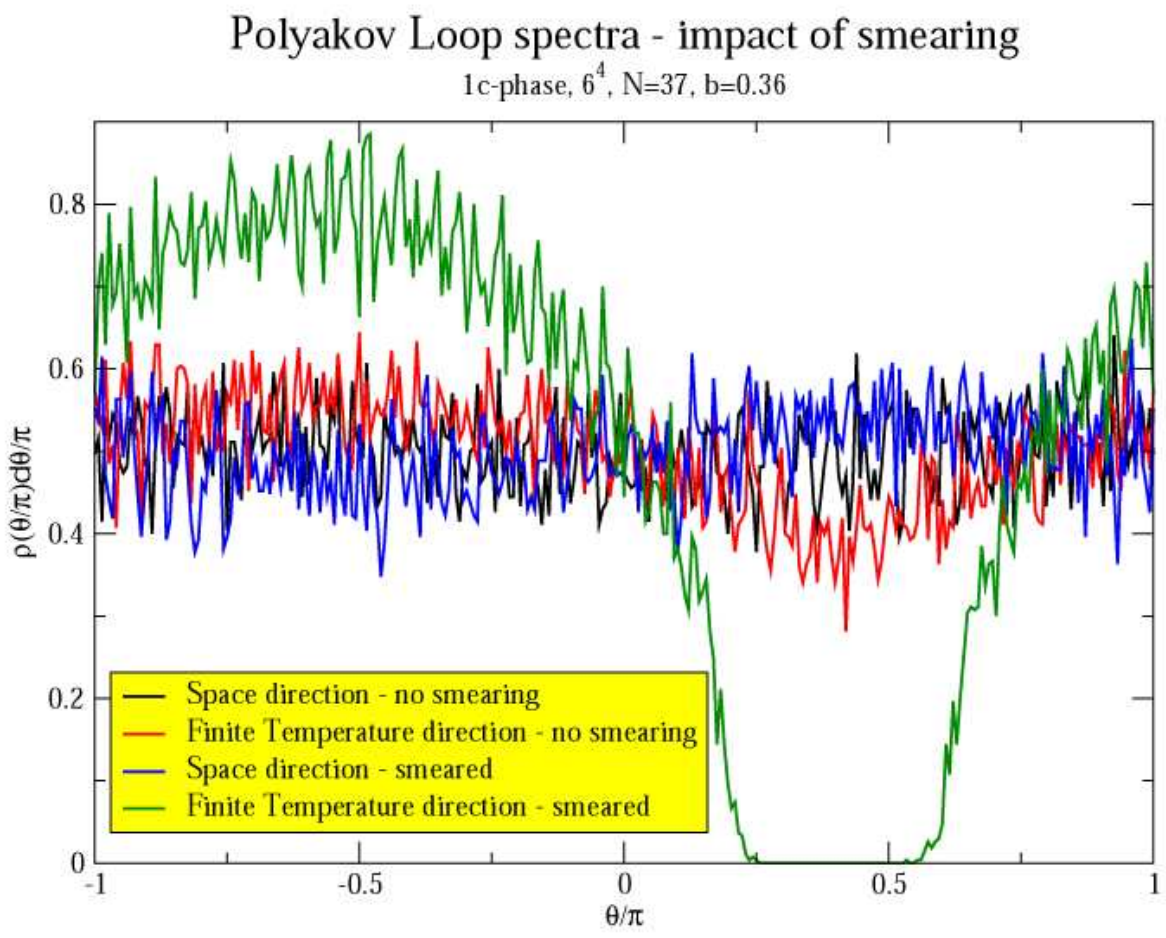

**Figure 4:** Spectra of a Polyakov loops with a gap and without in the 1c phase on a $6^4$ lattice at $N = 37$ and $b = 0.36$, spontaneously breaking the $Z(N)$ symmetry in one direction and preserving it in the other three.

The main characteristics of the phases are as follows:

- 0h is a “hot” lattice phase. Full reduction holds: traces of Wilson loops are independent of $L$ at infinite $N$. The open loop round a plaquette has no gap in its spectrum, and consequently the space of gauge fields is connected. In the continuum, this space is disconnected.

- 0c is the first [as $b \equiv 1/\lambda$('t Hooft) increases] continuum phase. In it the traces of all infinite $N$ Wilson loops still are independent of the physical torus size. Open loops associated with single plaquettes have a substantial gap and for large enough but finite $L$ and $b$, the space of gauge fields dynamically splits into disconnected components. In other words, with proba-bility one, at infinite $N$, all plaquettes are very close to unity in norm. This is very useful for simulating overlap fermions [10, 11]: the gap in the plaquettes makes the kernel of the sign function in the overlap operator have no modes close to zero that need to be projected out.

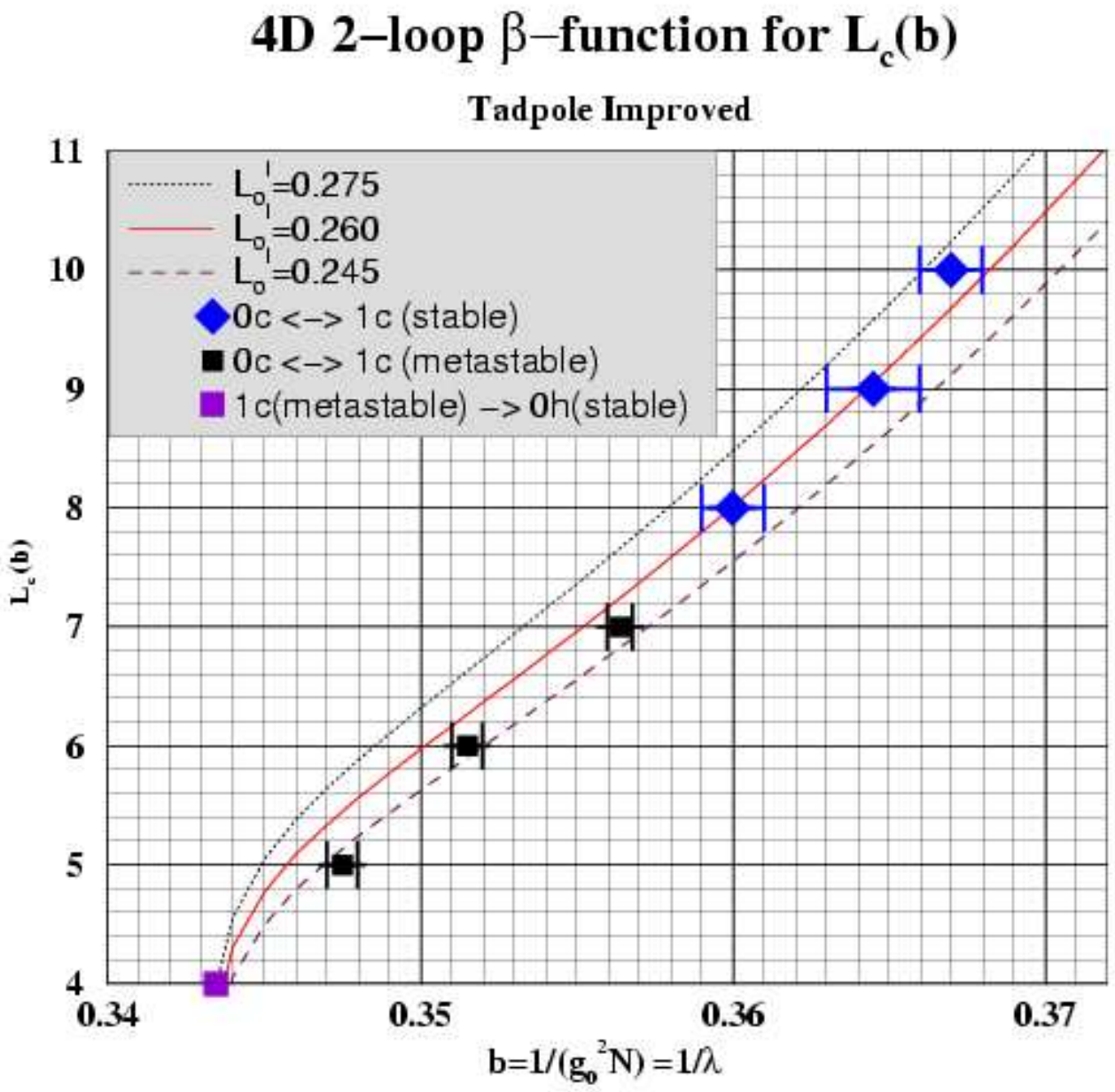

**Figure 5:** The 0c→1c transition points.

- In Xc=[1-4]c the $Z(N)$'s in X directions break spontaneously, and independence of the size in the corresponding directions is lost, but preserved in the other directions. Hence, 1c represents finite temperature planar, deconfined, QCD.

The phase 0c is of main interest as it corresponds to planar QCD at infinite Euclidean volume. The 0h→0c transition is well understood from earlier work [13], and for $L > 8$ occurs at a fixed $b$, $b \approx 0.36$ (as already mentioned, below this $b$ 0c can be maintained by metastability). The first objective is to study the 0c→1c transition. It is suspected to be the physical deconfinement transition and therefore ought to show the right scaling dependence of the critical size $L_c$ on $b$; since $L$ is an integer it is better to think about the inverse function, $b_c(L)$. True asymptotic scaling occurs at prohibitively large $b$ with the simple Wilson action, but the problem can be significantly remedied using tadpole improvement. In this way the data in the Figure 5 is brought in agreement with asymptotic freedom, making our numerical case for the physical nature of the 0c→1c transition.

The plot of $L_c$ as a function of b is well described by a tadpole improved formula:

$$b_I = b\frac{b^2-0.58960b+0.08467}{b^2-0.50227b+0.05479} \approx b - b_o + \mathcal{O}\left(\frac{1}{b}\right)$$
$$L_c(b) = (0.260 \pm 0.015)\left(\frac{11}{48\pi^2 b_I}\right)^{\frac{51}{121}} e^{\frac{24\pi^2 b_I}{11}}$$

## 6. Spontaneous chiral symmetry breaking

While reduction holds for pure gauge observables, the situation for fermionic observables needs re-thinking. One intuitive view of reduction is that an object in the fundamental representation will emerge from the propagation around compactified dimension of the torus with an orientation in color space that is orthogonal to that it held in a previous passage and so consecutive windings add up incoherently. As a result, a Wilson loop which is many times folded over itself on a small torus behaves the same as if it were stretched out in infinite space, with no folds. If our observable is a color singlet fermion bilinear, when it moves round a compact directions, there is no mechanism for incoherence. However, this observable can be thought of as a local creation operator for a meson and, while meson propagation certainly is sensitive to the compactification, mesons do not interact at infinite $N$, so the volume dependence likely is trivial. It turns out that the condensate associated with spontaneous chiral symmetry breaking has an infinite volume limit at large $N$ exactly determined by data on finite lattices. After all, reduction turns out to apply also to the condensate. At infinite volume and at infinite $N$ the condensate is given by the the infinite $N$ limit taken at some fixed finite volume. An infinite volume limit is unnecessary.

This works because S$\chi$SB at $N=\infty$ is described by the random matrix model (RMT) of Shuryak and Verbaarschot [14]. The overlap fermion propagator [11, 12] on valence fermion lines is given by the matrix $A$, which in the chiral basis has a block structure:

$$\gamma_5 = \begin{pmatrix} 1 & 0 \\ 0 & -1 \end{pmatrix},\; A^{-1} = \frac{1-\gamma_5\varepsilon(H_W)}{1+\gamma_5\varepsilon(H_W)} \to A = \begin{pmatrix} 0 & C \\ -C^\dagger & 0 \end{pmatrix}$$

Here $H_W$ denotes the Wilson Dirac operator at negative Wilson mass of order $-1$.

Shuryak and Verbaarschot [14] make C random, with an enhanced symmetry, much in the spirit of Wigner's approach to complex nuclei:

$$p(C)d^{2n^2}C \propto e^{-\kappa^2 n \mathrm{Tr}(C^\dagger C)} d^{2n^2}C$$

The model is usually framed in a continuum language, but the presence of an ultraviolet cutoff is immaterial so long as the needed symmetries are preserved. One does not need Lorentz invariance. The RMT model applies directly on the lattice and it is an unnecessary complication to also take the continuum limit at the stage that RMT is employed.

The size of $C$, $n$, is $2^{\frac{d}{2}-1}L^dN$, where $d$ is the (even) dimension. Just like in Wigner's case, RMT holds when $n$ is large enough. It is immaterial whether $n$ is made large by increasing $L$ or by increasing $N$. The dependence on $L$ becomes trivial as a result of taking $N$ to infinity, reflecting reduction. As $n$ increases, more and more of the very low eigenvalues start being described by RMT. This description has only one free overall parameter, determining a common scale for all the eigenvalues. All ratios among eigenvalues are random variables whose distribution is completely specified. We increase $N$ until the ratio of the smallest to the next smallest eigenvalue obeys the RMT universal function. Then, fits to the individual eigenvalue distributions give consistent estimates for the chiral condensate.

In Figure 6 several histograms of the two lowest eigenvalues of $\sqrt{-A^2}$ are shown. All the data has been collected on an $6^4$ lattice with $N$ increasing from 13 to 43, over various prime numbers.

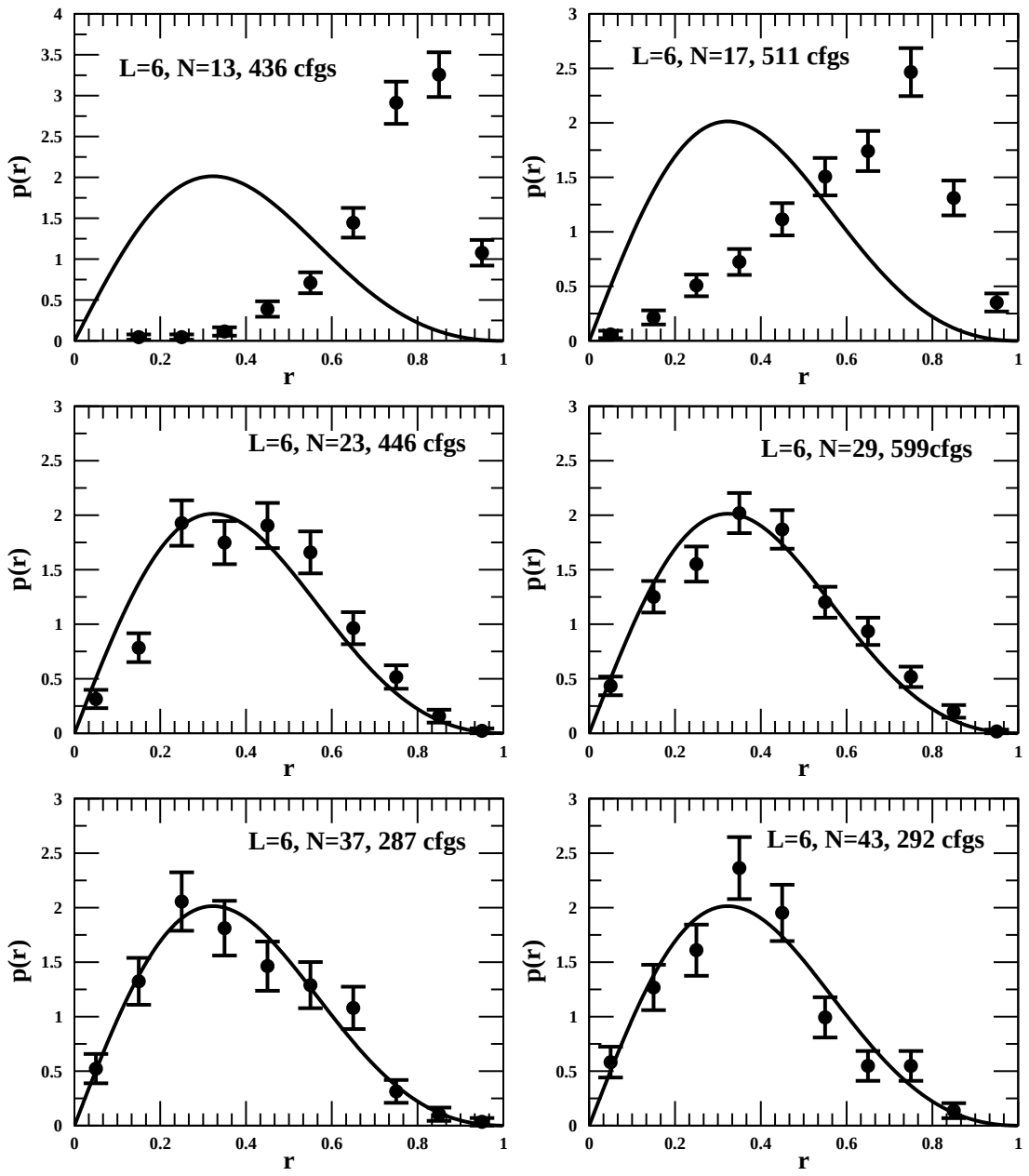

**Figure 6:** Convergence of the eigenvalue ratio to a universal, parameter free curve.

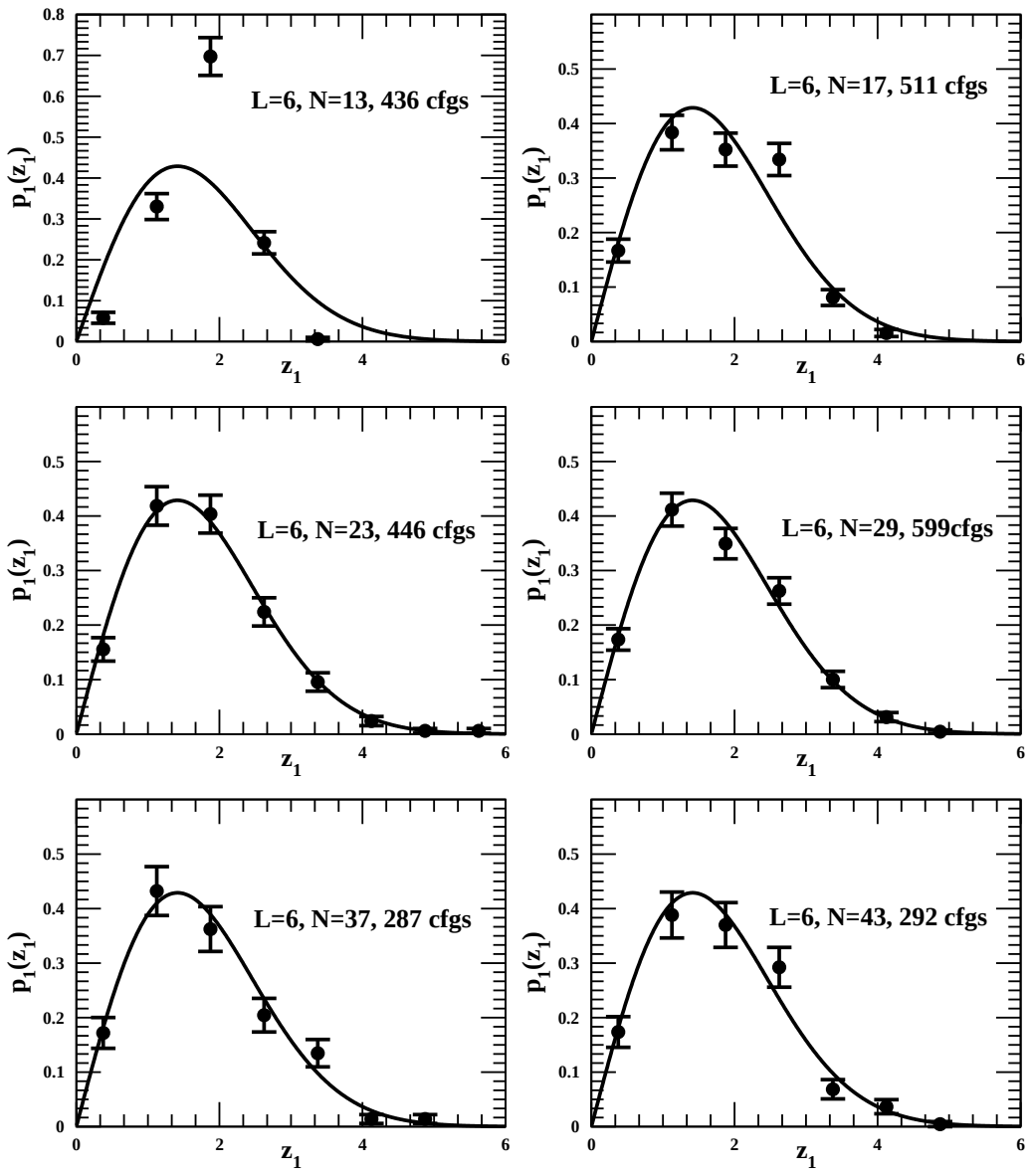

**Figure 7:** Determination of the chiral condensate from the smallest eigenvalue.

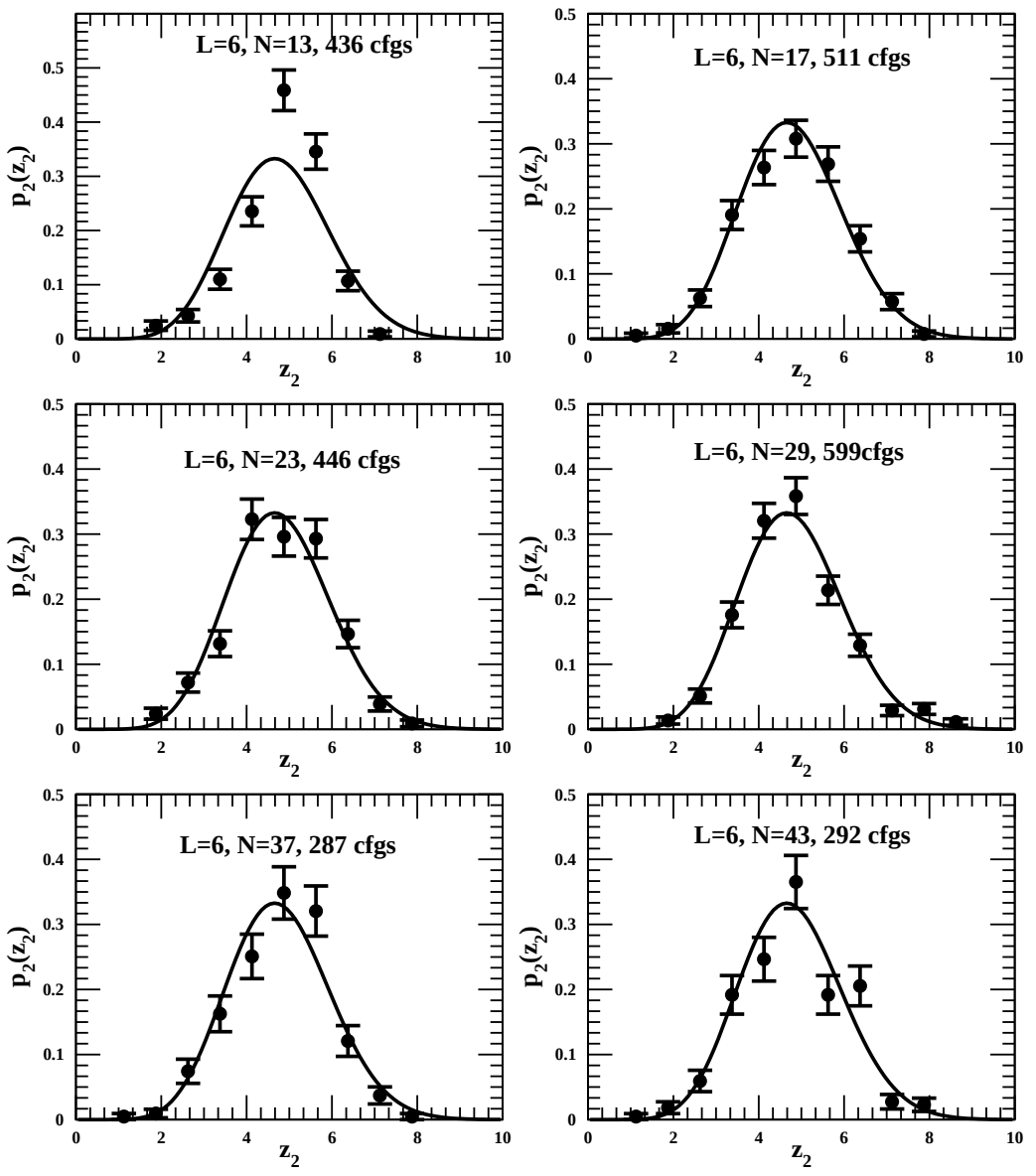

**Figure 8:** Determination of the chiral condensate from the second smallest eigenvalue.

We see how the RMT large $n$ limit is approached as $N$ increases. For $n$ too small, eigenvalue repulsion is not strong enough relatively to the other forces governing the eigenvalues and the ratio has a tendency to accumulate near its upper bound, unity. As $n$ increases, more and more eigenvalues have to fit into a constant interval, and repulsion becomes important. Eventually, the universal RMT behavior is attained. We verified numerically that as $L$ is increased, the value of $N$ where RMT behavior sets in gets lower. Once the RMT regime is entered one can use the individual eigenvalue distributions to get fits to the free parameter that translates into the chiral condensate. Only the two lowest eigenvalues have to be calculated for each gauge configuration. To apply the method one needs relatively high statistics, of the order of at least few hundreds of gauge configurations, if one wants to obtain a histogram approximating well the theoretical distribution. Fits to the lowest and next lowest eigenvalues are shown in Figures 7 and 8.

In conclusion, at $b = 0.35$, in a metastable extension of the 0c phase, all the data with $N_c \geq 23$ fits RMT with a chiral condensate given by $\Sigma \sim (0.14)^3$ in lattice units. Varying the coupling $b$ and repeating the procedure one can eventually translate the lattice results into continuum ones. The definition of the condensate is ambiguous because of the non-zero anomalous dimension of the associated operator. Using a perturbative tadpole-improved estimate for the $Z_S$ normalization constant in the $\overline{MS}$ scheme we obtained:

$$\lim_{N\rightarrow infty} \frac{1}{N} \langle \bar{\psi}\psi \rangle^{\overline{MS}} (2\ GeV) \approx (0.65 T_c)^3 \approx (174\ MeV)^3$$

To the extend one can attach a meaning to corresponding estimates obtained at $N = 3$, the result indicates a small $\frac{1}{N}$ correction, even for $N = 3$. The standard value for the infinite $N$ finite temperature deconfinement transition, $T_c$, is used to set the scale in the above equation.

## 7. Reduction for Mesons: $L$ independence.

As explained in the previous section, reduction needs to be carefully thought out when applied to specific mesonic observables. Here we turn to the question of meson masses. This touches upon one of the most important long term objectives of any study of planar QCD, namely obtaining the asymptotic behavior of heavy meson resonances and seeing their organization into Regge trajectories. As even the briefest review of string history reveals, it is of paramount importance to first make sure that indeed one has massless pions when the quark mass vanishes. Historically, this fixed the intercept of the $\rho$ meson trajectory. Although we have established spontaneous symmetry breakdown in the planar limit, the pion mass is not an entirely trivial matter, since chiral symmetry breaks at infinite $N$ already at finite volume, and there are no continuous momenta at finite volume, so Goldstone's theorem does not extend in a straightforward manner.

The way out is to "force feed" momentum [15] into the quark line description of the pion (or any other meson for that matter). This procedure is exact at infinite $N$, and lets one evaluate the pion propagator at any momentum on a lattice as small as $6^4$ ! An alternative view of the procedure is to say that the quark boundary conditions are given flavor dependent "twists". However, in the large $N$ literature the term "twist" is already reserved for something else, that will be addressed later on, and therefore we shall stick to the older term of momentum "force feeding".

Let us take two degenerate fermion flavors. The following expressions provide interpolating fields for flavor non-singlet, gauge singlet mesons:

$$M(x) = \frac{1}{\sqrt{N}}\bar{\psi}(x)\chi(x), \quad \bar{M}(x) = \frac{1}{\sqrt{N}}\bar{\chi}(x)\psi(x)$$

We are interested in the Fourier transform of the flavor non-singlet meson:

$$\tilde{S}(x,y) = \langle \bar{M}(x)M(y)\rangle$$

The quark propagator is dependent on the gauge background $U$:

$$\langle \psi(x)\bar{\psi}(y)\rangle_{\{U\}} = G(x,y,\{U\})$$

We now introduce an associated gauge field background $U^{(q)}$:

$$q_\mu = 1\cdots N_c; \;\; k_\mu = 1\cdots L; \;\; Q_\mu = \frac{2\pi}{LN_c}q_\mu; \;\; K_\mu = \frac{2\pi}{L}k_\mu; \;\; U^{(q)}_\mu(x) = e^{\imath Q_\mu}U_\mu(x)$$

We define the meson propagator in Fourier space, for arbitrary (at infinite $N$) momentum $Q+K$:

$$S(Q+K) = -\frac{1}{N_c}\sum_x e^{\imath K\cdot(x-y)}\langle Tr[G(x,y,\{U^{(q)}\})G(y,x,\{U^{(0)}\})]\rangle$$

The above propagator is evaluated in a Monte Carlo simulation with the help of Gaussian noise vectors. More precisely, the propagator is evaluated for a pre-chosen set of quark masses, where the same gauge background and noise vectors are used for all masses. This makes the overhead from including several quark masses negligible at the expense of high correlations between the estimates at each quark-mass. This time one needs to pay attention to the order of the limits $N\to\infty$ and $m_q\to\infty$, as discussed earlier. (In the case of the condensate we succeeded going around this obstacle, courtesy of RMT.)

## 8. Pion mass versus quark mass [9].

While it is true that we can now work in Fourier space directly, and look directly for poles in the propagator, there is nothing to protect us from higher mass states contributing to the same field theoretical correlation function. The technique we used for the pion case relies heavily on the theoretical conjecture that at high masses chiral symmetry is effectively restored and meson masses come in parity doublets. Phenomenological studies indicate that in QCD this “restoration” effectively takes place at relatively low energies [16]. Obviously, at very low masses there is no such parity doublet degeneracy and the pseudoscalar pions are singled out as massless. Thus, by looking at a linear combination of correlators to which parity doublets make contributions that cancel out, we can hope to vastly diminish the contribution of most heavy masses. To further reduce the contribution of the nearby higher states we use a smeared version of the fermion bilinear representation of the meson field where smearing is affected by using the inverse covariant lattice Laplacian.

The data so obtained is fit to extract the pion mass pole as a function of quark mass in lattice units. The error correlation matrix in the space of quark masses is badly conditioned, reflective of the fact that one does not have as much information as the number of different quark masses might indicate. We chose to treat this problem by a PCA (principal component analysis) method, in which we only kept a few linear combinations over quark masses. The selected combinations correspond to the largest eigenvalues of the correlation matrix included until they saturate about ninety percent of its trace. After this pruning, the data produces reasonable error estimates, which are ultimately obtained using single elimination jack-knife bootstrap.

The data was fit to a single pole term. This procedure was tested against fits that allowed for a free additive constant (a background term). The background term proved to be irrelevant. In two dimensions, where the effective “restoration” of chiral symmetry is known not to hold, a background term is seen numerically to be necessary. Thus, our analysis provides indirect support for the conjecture of effective chiral symmetry restoration at high energies in the planar limit. Theoretically, the arguments for this conjecture seem to be cleanest at infinite $N$.

The above procedure produces pion masses in lattice units. It is of interest to see the result as determining the pion mass as a function of quark mass. The latter is multiplicatively ambiguous in the continuum. We replaced the quark mass by the product $m_q\Sigma$ where $\Sigma$ is the chiral condensate at zero quark mass and same lattice spacing. This product is unambiguous. We use the deconfinement temperature, $T_c$ to set the scale, converting the masses into physically unitless quantities.

To analyze the data one needs some parameterization of the quark mass dependence of the pion mass. At small quark masses the parameterization of choice is that given by the chiral Lagrangian. At infinite $N$ chiral Lagrangians are less informative than at finite $N$. The pions are free massless particles, so the softness of their scatterings at small momenta is a side comment on the total suppression of these interactions by $N$. Little is left of the chiral Lagrangian machinery beyond the pion mass dependence on quark mass. For the latter we employed a parameterization that holds the promise to apply in a larger range of quark masses, while preserving agreement with ordinary chiral Ward identities. This parameterization is motivated by gravitational duals obtained using string theory and by a direct solution of the two dimensional ’t Hooft model.

Here are some details about the ’t Hooft model: The meson spectrum is determined by an equation that sums up all planar diagram contributions to the poles in the propagator of a meson composite field. That equation has the following form, known as 't Hooft's equation:

$$\gamma\left(\frac{1}{x}+\frac{1}{1-x}\right)\phi(x)-P\int_0^1\frac{\phi(y)-\phi(x)}{(y-x)^2}dy=\mu^2\phi(x)$$

$x$ is the fraction of meson light-cone momentum carried by one of the quarks and varies between 0 and 1. The gauge coupling is used to set the scale and the dimensionless meson mass squared is $\mu^2$ while the dimensionless quark mass squared is $\gamma$. Some changes of variables facilitate an expansion around $\gamma=0$; this expansion is a derivative expansion, but not a chiral expansion. At leading order, it relates $\mu^2$ to $\gamma$ by a Schrödinger equation:

$$\frac{\pi^2}{3}\frac{d^2}{ds^2}\Psi(s)+\frac{\mu^2}{4\cosh^2\frac{s}{2}}\Psi(s)=\gamma\Psi(s)$$

Here, $\Psi(s)$ is the meson wave function and $s$ is a rapidity variable. The decomposition of this equation into $-R^\dagger R$+Const. , with $R$ of first order in $\frac{d}{ds}$, induces one to replace $\mu^2$ by a variable $\Delta$:

$$R=-\frac{\pi}{\sqrt{3}}\frac{d}{ds}+\Delta\tanh\frac{s}{2};\;\; \frac{1}{4}\mu^2=\Delta(\Delta+\frac{\pi}{2\sqrt{3}});\;\; \Delta=\frac{1}{2}\left[\sqrt{\mu^2+\left(\frac{\pi}{2\sqrt{3}}\right)^2}-\frac{\pi}{2\sqrt{3}}\right]$$

It is more natural to expand $\Delta$ in $\sqrt{\gamma}$ than to expand $\mu^2$ in $\sqrt{\gamma}$; one series is a rearrangement of the other. At leading order, $\Delta=\sqrt{\gamma}$. From the series giving $\Delta$ in terms of $\gamma$, and starting with the Schrödinger equation above, the complete equation of 't Hooft can be reconstructed, order by order in the derivative expansion.

In four dimensions we were thus lead to parameterize the data by

$$\Delta=\frac{1}{2}[\sqrt{\frac{m_\pi^2}{T_c^2}+\Lambda_\pi^2}-\Lambda_\pi];\;\; \Delta=\frac{m_q\Sigma}{T_c^4}+\frac{m_q^2\Sigma^2}{T_c^8}\frac{1}{\Lambda_q}+\ldots,$$

where the scale is set by:

$$T_c=\frac{1}{l_c}\approx 264\ MeV$$

As $m_\pi\to 0$, $\Delta\propto m_\pi^2$; it is $\Delta$ that is fit to a series in the quark mass, not $m_\pi^2$. This entire discussion is predicated on the absence of chiral logs in the planar limit. The resulting fits are shown in Figure 9. The fits also produce the pion decay constant at infinite $N$:

$$\frac{F_\pi}{\sqrt{N}}\equiv f_\pi=\frac{T_c}{\sqrt{2\Lambda_\pi}}\approx 71MeV$$

Going back to $N=3$ we can compare to real QCD:

$$N=3\to F_\pi=123MeV$$

The experiment based value for real QCD is $F_\pi=86\ MeV$ at zero quark mass.

In summary, we described the data using a string motivated parameterization:

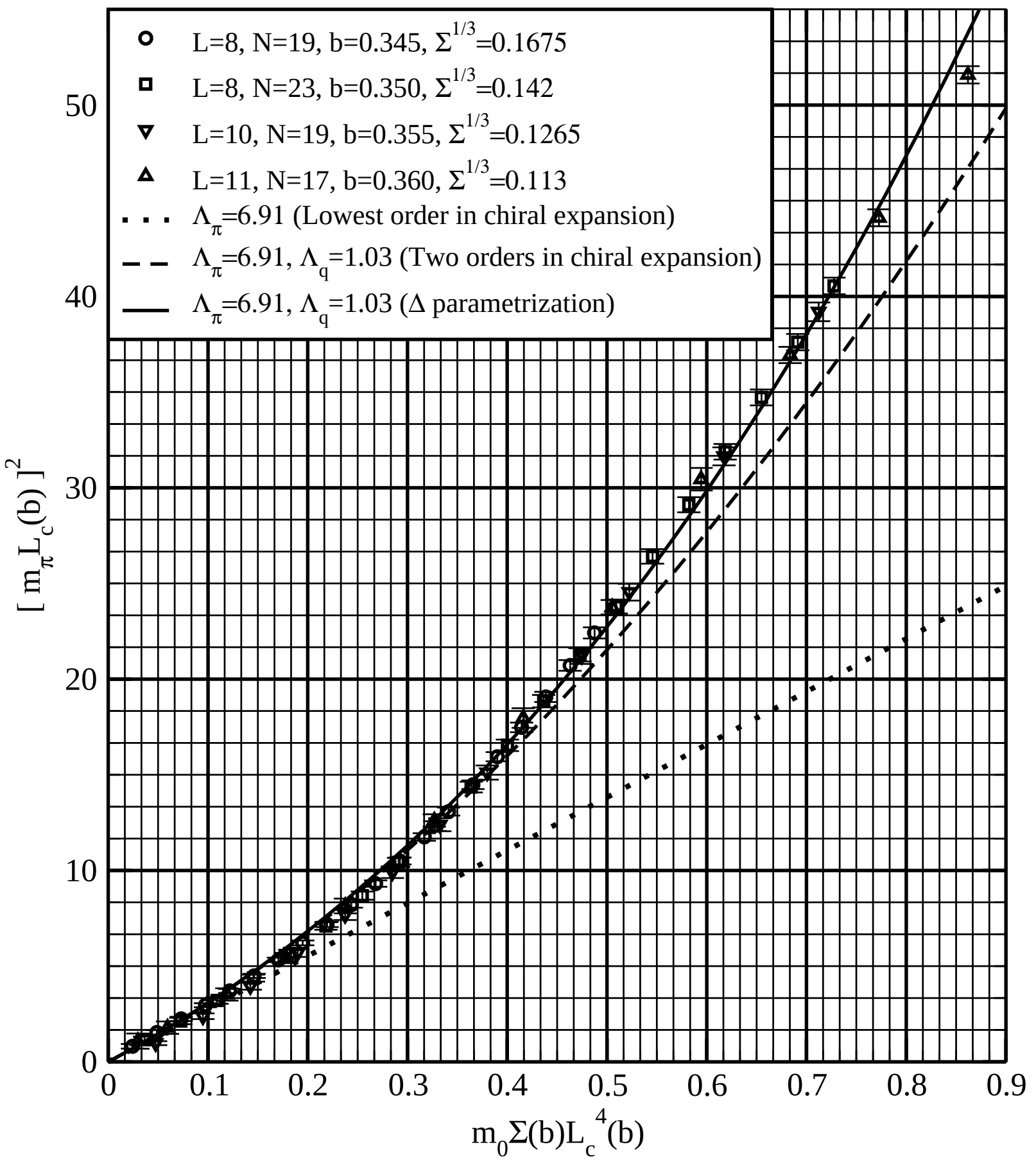

**Figure 9:** Pion mass squared as a function of quark mass.

- The parameterization in terms of $\Delta$ is reminiscent of mass formulae in cases where the AdS/CFT correspondence holds and of explicit formulae in planar 2D QCD.
- In principle, $m_\pi^2(m_q)$ contains enough information to determine the warp factor in a hypothetical 5D string background, if this were all that is needed to define the string dual of QCD. Explicit manipulations of planar 2D QCD provide an analogue to this conjectured role of $m_\pi^2(m_q)$. A plot of our four dimensional results is shown in Figure 9.

## 9. The 1c phase

We now turn to the 1c phase including a report on preliminary results. Let us first summarize the situation:

- The 0c phase corresponds to infinite volume planar QCD.

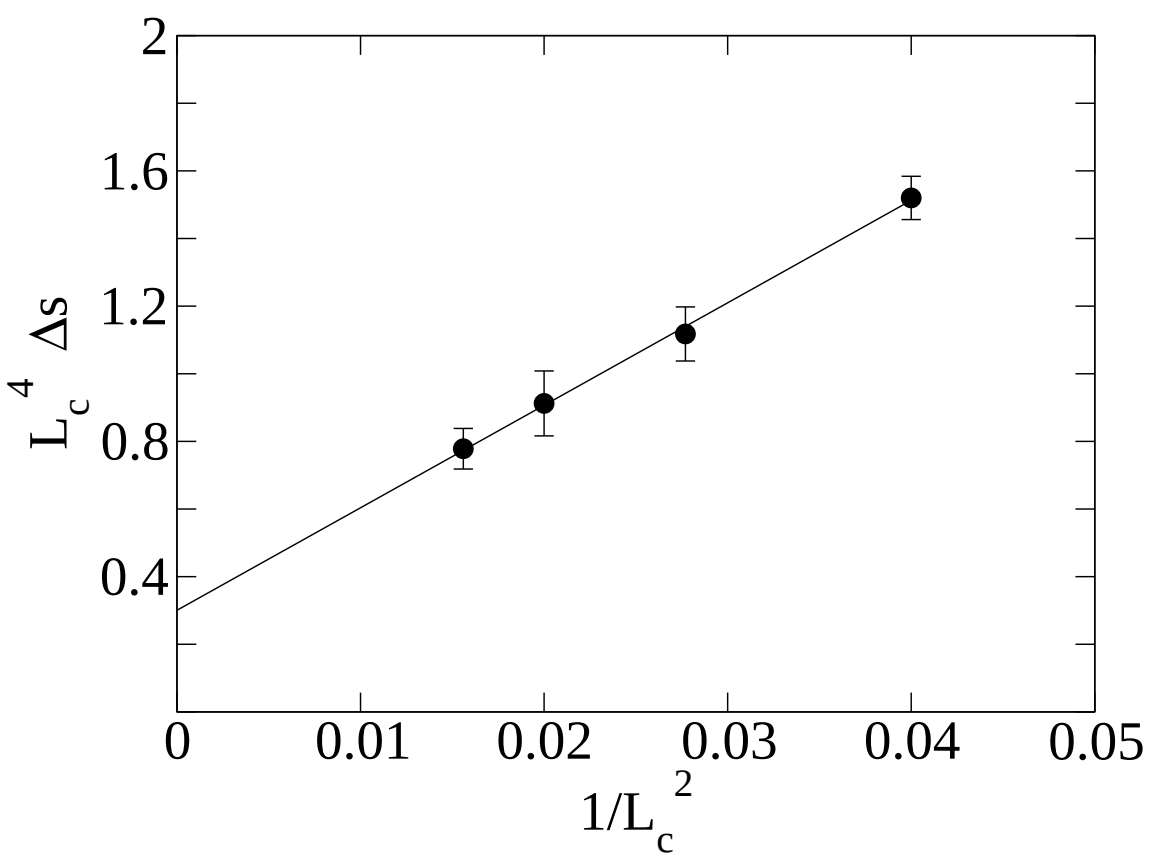

**Figure 10:** 0c → 1c scaling plot of jump in action across the 0c→1c boundary, from [17].

- In 1c, one direction is selected dynamically to play the role of a finite temperature direction. Note that the system can sustain only a high enough temperature. As in string theory, the free energy at leading order is temperature independent so long as the temperature is low enough.

- The following tests were carried out to show that 1c indeed is planar QCD at infinite space volume and finite temperature: (a) Kiskis [17] has checked the scaling of the latent heat, (b) We have obtained preliminary evidence that chiral symmetry is restored in the 1c phase where the Euclidean massless Dirac operator develops a spectral gap approximately linear in temperature.

Figure 10 is reproduced from [17]: $\Delta s$ is the jump in action at the 0c → 1c transition point and $L_c(b) = \frac{l_c}{a}$ is used to monitor scaling violations. The data is consistent with a dominating scaling violating term of order $a^2$, where $a$ is the lattice spacing. The extrapolated number is in rough agreement with results of Teper et. al.

We now turn to some preliminary data on chiral and axial-$U(1)$ symmetry restoration at finite temperature. Two points are basic:

- At infinite $N$ the finite temperature gauge transition drags the fermions to restore chiral symmetry.

- For $T > T_c$ the massless overlap Dirac operator develops a spectral gap at zero; the axial $U(1)$ symmetry is therefore restored as well.

The massless Euclidean Dirac operator in a special finite temperature background which has the zero component of the gauge field set to zero, would have a gap proportional to the temperature in its spectrum for arbitrary static spatial gauge fields. The chiral condensate would be zero and chiral symmetry would be restored. It is expected that for high enough temperature chiral symmetry will be restored for all typical gauge field backgrounds. A priori, the temperature at which chiral symmetry is restored does not have to be the same as the deconfinement temperature in the quenched approximation.

At infinite $N$ the deconfinement temperature $T_c$ is a symmetry breaking point determined by the gauge fields since the fermions are automatically quenched. For $T > T_c$ there is a set of $N$ degenerate vacua and it is not justified to keep on neglecting the fermions as they lift the degeneracy among the different vacua at order $N$. The Dirac operator is defined with anti-periodic boundary conditions in the temperature direction; the boundary conditions in the spatial directions do not matter, since the associated $Z^3(N)$ global symmetry group stays unbroken in 1c. As the various gauge vacua are considered in turn, the one with the largest determinant will dominate. The various gauge vacua can be parameterized by one angle $\theta$ which is the average phase of the Polyakov loop in the temperature direction. Figure 11 shows the dependence of the gap on $\theta$ via the expectation value of the smallest eigenvalue of $\sqrt{-A^2}$. CP invariance and common sense imply that the chosen gauge vacuum will be the one that has a positive trace for the temperature Polyakov loop. Figure 11 shows that the largest spectral gap is obtained at the smallest $\theta$, and it is plausible that when the spectral gap in the Dirac operator is largest, the fermion determinant itself is maximized. The gap is given by $\pi T(1-\frac{\theta}{\pi})$ for free fields and the linear dependence in Figure 11 indicates that the only visible effect interactions have is an effective reduction in the temperature from the free field value of $\frac{1}{L_4}$. With this understanding about the choice of vacuum, the fermions can be taken as quenched to leading order in $\frac{1}{N}$, even in 1c. In practice the fermionic boundary conditions are adjusted so as to be antiperiodic with respect to the phase of the trace of the temperature Polyakov loop. The particular vacuum that is randomly selected by the gauge fields in 1c is used to determine the physically correct boundary conditions for the fermions and with that done one can continue to ignore the fermion loop contribution to the gauge field probability distribution.

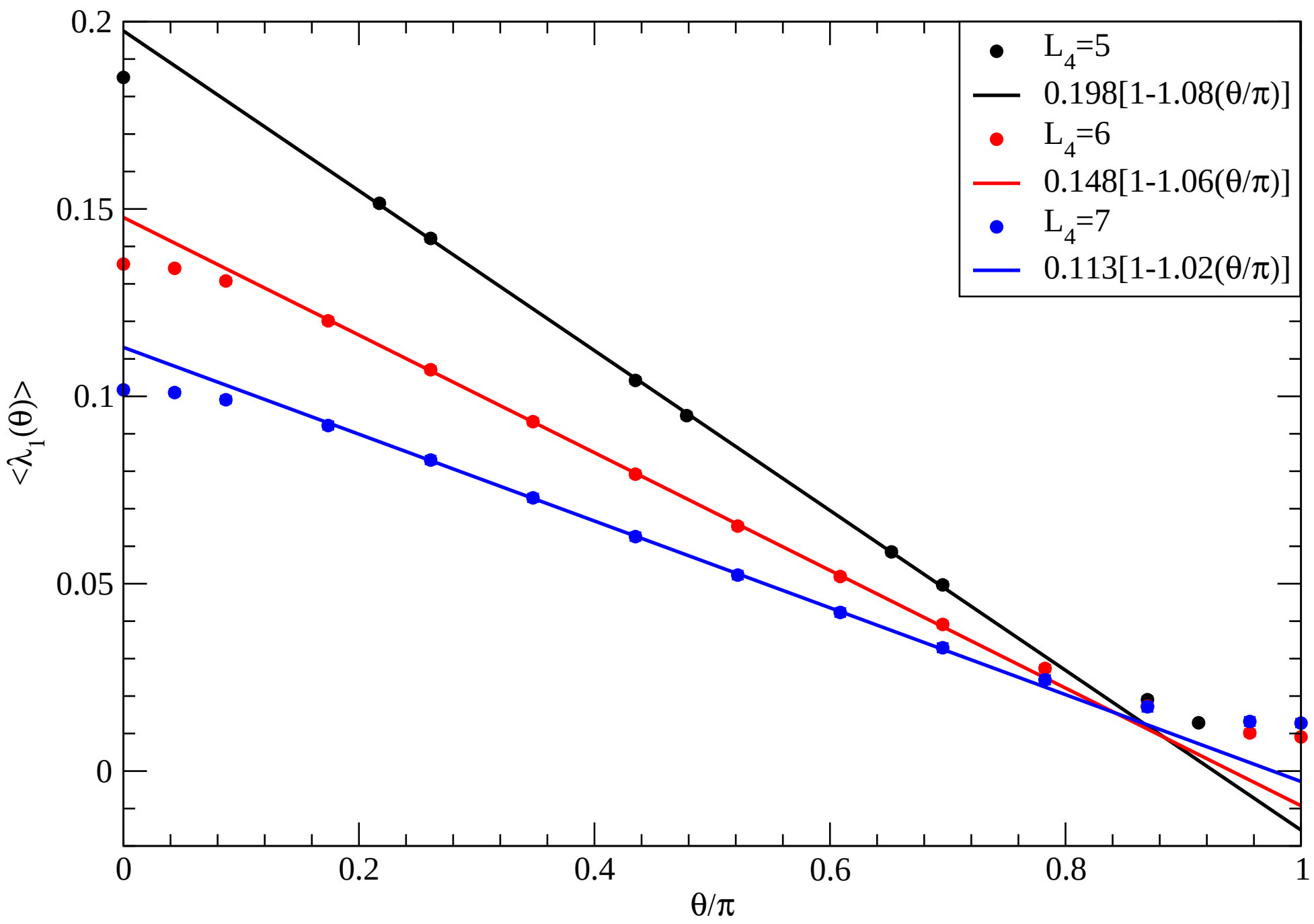

**Figure 11:** Dependence of the gap in the overlap Dirac operator as a function of fermion boundary conditions in the 1c phase with $b = 0.36$ on lattices of size $8^4 \times L_4$.

The simplest view of the deconfined phase has the Polyakov loop operator proportional to a unit matrix, but we know now that this is untrue at infinite $N$. Rather, it has a spectrum with a support extending over a fraction of the circle. If the fermions of different colors were decoupled, only some would have approximately antiperiodic boundary conditions; others, would come in pairs experiencing complex conjugate twisted boundary conditions with various phases. Consider a single fermion in a spatially static background with $A_0 = 0$ and twisted boundary conditions: As the twist goes from -1, corresponding to antiperiodic boundary conditions to +1, corresponding to bosonic boundary conditions, while the determinant stays at power +1, because of statistics, the contribution to the pressure changes sign. Indeed, the power of the determinant would have to also change sign as the boundary condition does if the contribution to the pressure were to be kept positive. Thus, we expect the width of the spectrum of the Polyakov loop operator to reduce the pressure contribution of the fermions from its classical value. The fermions are massless, so the contribution must go as $T^4$ asymptotically, and only its coefficient is expected to be effectively reduced at temperatures not too far above $T_c$. Eventually, as $T$ goes to infinity, the width of the Polyakov loop spectrum shrinks to zero and the classical result for the fermion pressure will be approached. It is unreasonable to expect the different fermion colors to be even approximately decoupled in real 1c gauge backgrounds. If this were the case, the gap would close in Figure 11 for some $0 < \theta^* < \pi$ where $\theta^*$ gets close to 0 as $L_4$ increases. This is not the case and one has a gap for all values of $\theta$. Yet, Figure 11 does indicate that a remnant of the above mechanism survives, and before the regime of aymptotically high temperatures is entered, the coefficient of the $T^4$ term in the pressure is effectively averaged down by an amount decreasing as $T$ increases. The intuitive picture is that the coupling among colors in combination with the spread of Polyakov loop eigenvalues induces an effective averaging over fermionic boundary conditions that is symmetric around the antiperiodic choice. Even though the spectrum of Polyakov loop operators typically covers a large fraction of the unit circle, the Dirac operator with the right boundary conditions always has a gap in 1c. The averaging simply diminishes that gap relatively to the expression that holds at very high temperature.

In the static case, the dependence of the gap on temperature is determined trivially. The gap is linear in temperature for high enough temperatures; because of the spread of eigenvalues of the Polyakov loop, we expect the gap to approach its high temperature value from below as the temperature is increased. Since the Polyakov eigenvalues are relevant random variables influencing the gap, a simple isolated random matrix model description of the finite temperature spectrum of the Dirac operator in 1c is not expected to hold in spite of the non-sparse Dirac matrix becoming infinite in size as $N \to \infty$. Not only do we need to see the chiral random matrix model that held in 0c to stop applying in order for chiral symmetry to be restored, we hypothesize that no simple random matrix model can describe exactly even the low states in the spectrum of the Euclidean Dirac operator in 1c because the Polyakov loop variables play a special role. This is tested by looking at the pair correlation between the lowest eigenvalues of the Dirac operator. When normalized by the standard deviations of the individual eigenvalues that correlation tends to a number of order 1/3 to 1/2 in various random matrix models. It is bounded from above by unity, and we find that this bound is almost saturated in practice. This implies that there is a linear combination of the lowest eigenvalues that fluctuates substantially less than the typical fluctuations of the individual eigenvalues. This is not what simple random matrix models produce and indicates the presence of an additional scale governing eigenvalue fluctuations. Our tentative interpretation of the result is as a reflection of the relevance of the random dynamics associated with the eigenvalues of the Polyakov loop operator.

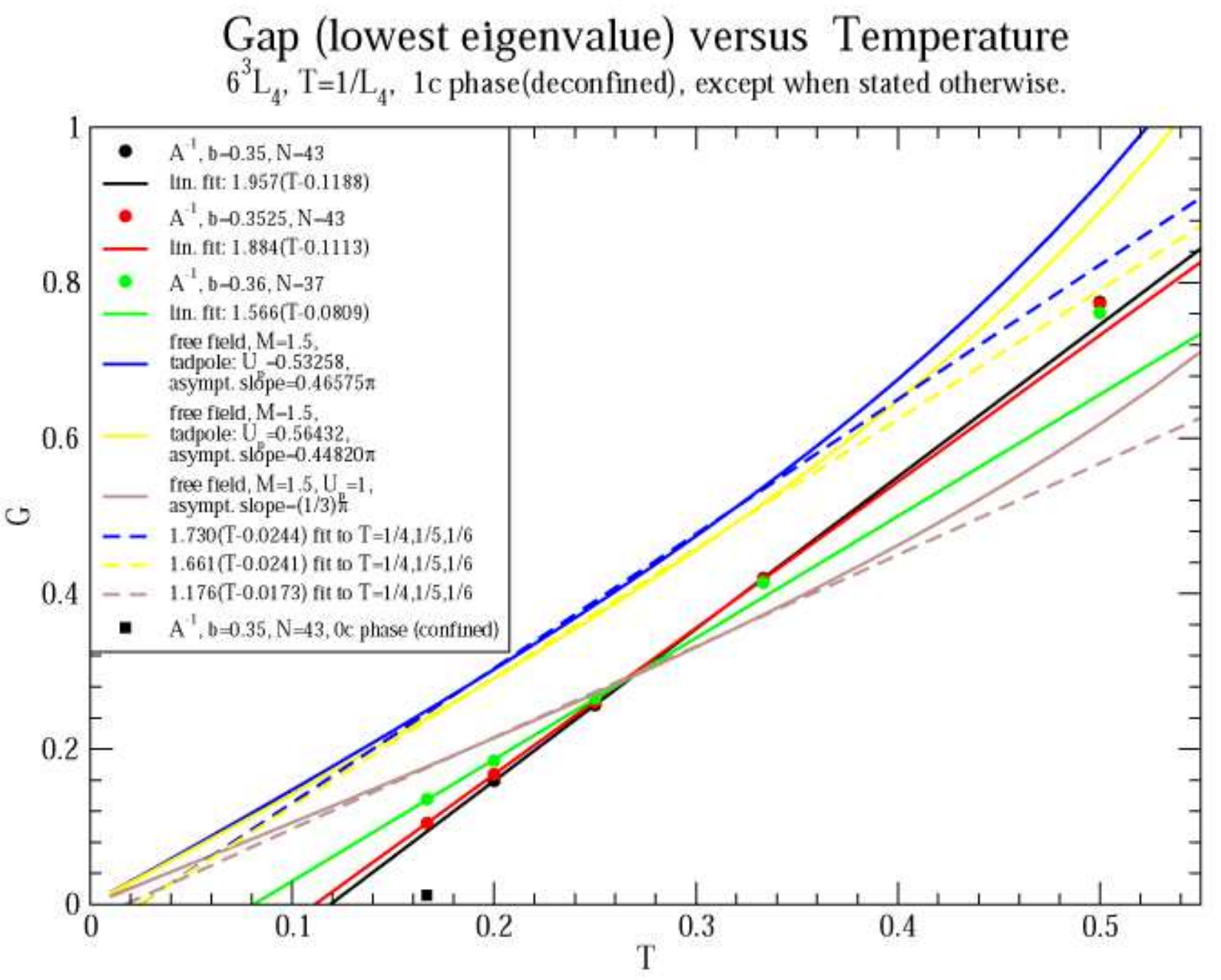

**Figure 12:** Study of the dependence of the gap in the overlap Dirac operator on temperature.

Our preliminary findings are consistent with our hypothesis, but we are limited to temperatures of the order of $T_c$ because of increasing lattice effects as $T$ increases.

In Figure 12 we show the averaged lowest eigenvalue of the Euclidean Dirac operator obtained on various lattices of size $6^3L_4$ at couplings $b = 0.35, 0.3525, 0.36$ in the 1c phase. We take the averaged lowest eigenvalue as a numerical definition of the gap. As $L_4$ varies at fixed $b$, $T_c$ stays fixed and $T$ varies. We vary $T_c$ by varying $b$; for $b = 0.35$, $T_c = 1/6$.

The data for $b = 0.35$ clearly shows that chiral symmetry is restored right at $T_c$, the deconfinement transition at infinite $N$. The gap is high and non-zero for $T = 1/5, 1/4, ..$ and drops to zero at $T = 1/6$ when the system goes over into the 0c phase. To be able to stay in 1c at $T = 1/6$ we need to increase $b$. We observe a linear dependence of the gap on $T$, as expected from the static picture. The linear dependence extends far into the high temperature domain, where lattice artifacts dominate; this was somewhat unexpected. This pattern holds also for higher values of $b$. The lines for different $b$'s intersect in some small region; only below that point can one expect any semblance to continuum physics. There we see that an increase in $b$ causes the gap to increase. An increase in $b$ makes $T_c$ go down in lattice units, and hence, if the temperature is held fixed in lattice units, we are at a higher $\frac{T}{T_c}$ ratio. We expect this to increase the gap towards the value it would have had if high temperature asymptotitcs applied. This is indeed what the data shows, so long as we stay at low enough temperatures, where lattice effects are reasonably small.

We compare our raw data to simple minded perturbation theory and to a tadpole improved version of this simple minded formula. The tadpole improved version produces lines parallel to the ones we obtained in our simulation but higher up. This is in agreement with the expectation that the gap will come out smaller than perturbatively expected. The most naive form of perturbation theory gives a different slope but goes through the region where all lines meet. This reflects an accidental balancing between different effects, some nonperturbative and some coming from ultraviolet fluctuations.

All our linear fits involve an intercept. The intercept can be thought of as the point where chiral symmetry would have broken spontaneously had we been able to super-cool the 1c phase to temperatures below $T_c$. It is hard to make too much of the intercept, because, as we show, fits of the perturbative results (tadpole improved and not) at the selected lattice temperatures also produce an effective intercept, albeit it substantially smaller than the one the real data gives. However, there clearly is no intercept in ordinary perturbation theory, so one cannot easily draw a quantitative conclusion about the chiral symmetry restoration temperature in the super-cooled deconfined phase. Moreover, it is unreasonable to expect the Dirac gap to stay strictly linear in temperature to the point where chiral symmetry would be spontaneously broken in a metastable, super-cooled, 1c phase. The symmetry breaking effect is a critical phenomenon in three effective dimensions and one ought to expect nontrivial critical exponents to be associated with it.

Obviously, this data is preliminary and the analysis we have presented could change as we make progress on this project.

## 10. Tricks with twists

There likely are further ways to reduce the amount of computational work needed to get the large $N$ limit of particular observables. One well known trick employs twisting the boundary conditions on the torus [18]. In general, the idea is that the twist has no impact on the results at infinite $N$ in the phase 0c, but the phase 1c must be replaced by something else, plausibly increasing the range in $b$ of the 0c phase at small $L$.

Below is a brief review of work being carried out in collaboration with A. González-Arroyo.

Twisted gauge fields can be defined over the continuum torus and fall into various distinct bundles of $SU(N)/Z(N)$ gauge theory. On the lattice the nontrivial boundary conditions can be absorbed into a phase of the coupling constant by a change of variables, thus making it explicit that the gauge field action is defined in a translational invariant way, or, in other words, that we indeed are working on a smooth torus. When we have twists, there are questions that arise about CP invariance and about the inclusion of fields in the fundamental representation, like fermions. For the simplest case of twist CP invariance is maintained, and the twist can be absorbed by introducing a limited amount of flavors and canceling the color twist against a flavor twist. We thus are led to a viable alternative for simulations intended at getting at the meson sector of planar QCD with non-dynamical fermions.

Just as in the untwisted case we have been looking at until now, there will be a critical maximal torus size, $L_c(b)$ for each $b$ beyond which the 0c phase cannot be maintained. Again, we expect $L_c(b)$ to grow asymptotically as a function of $b$ in a way compatible with asymptotic freedom. In practice we expect that twisting will make it possible to get to a given coupling $b$ (lattice spacing) with a smaller $L$ than in the untwisted case. How exactly this works out requires a numerical study.

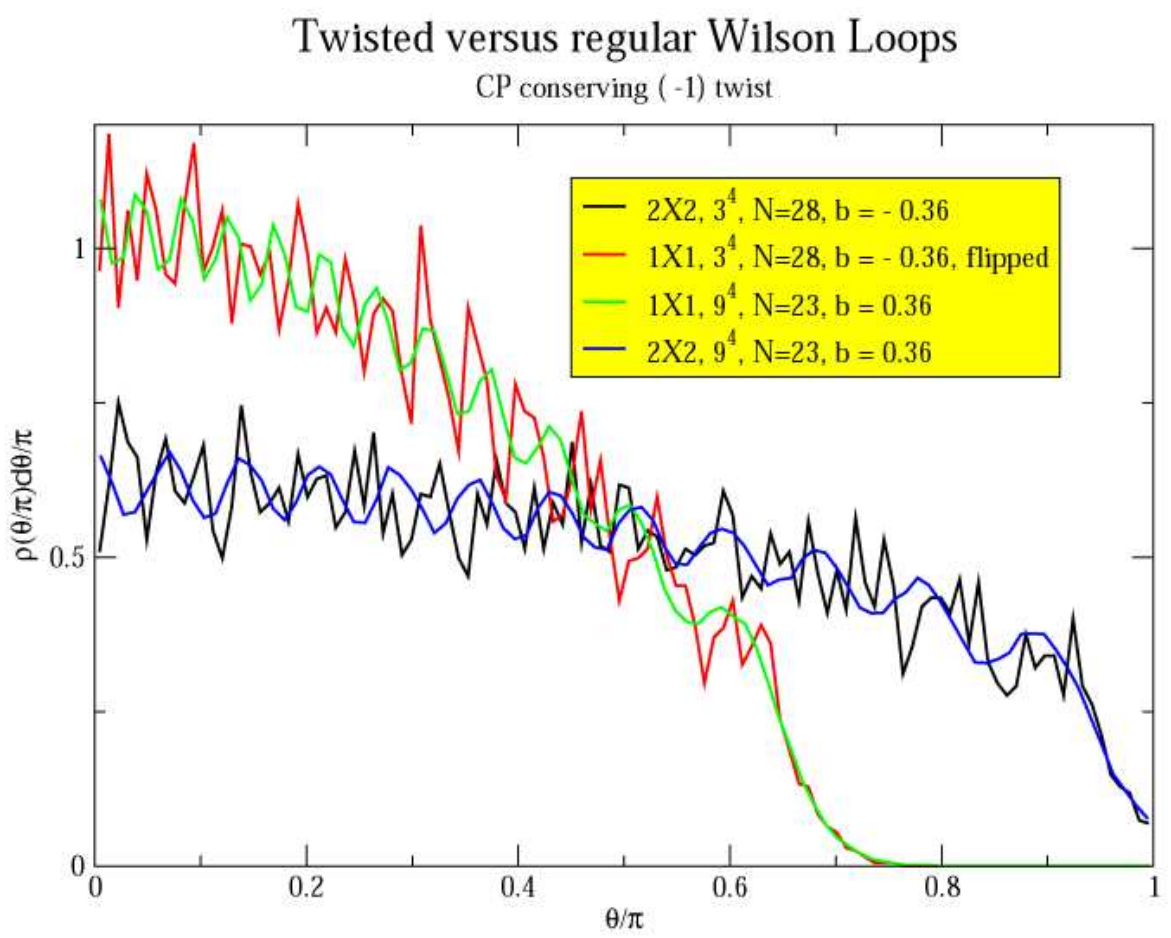

**Figure 13:** A comparison of twisted to untwisted results for two sizes of Wilson loops at b=0.36

One problem is that, unlike in the untwisted case, we do not have a clear physical picture for the phase 0c would decay into as the torus is shrunk. This makes it difficult to decide numerically whether or not the system still is in the 0c phase. Nevertheless, it is intuitively clear that one would expect 0c to extend to smaller physical volume sizes.

In summary, we want to exploit twists because:

- The 1c phase can be eliminated by minimal twisting: switch the sign of the lattice coupling $\beta$, take $L$ odd and $N = 4\times$(a prime).
- This twist preserves CP and extends the range of the 0c phase to smaller $L$'s at fixed $b = \beta/(2N^2)$.
- In the continuum, fermions can be added in flavor multiplets of 4; flavor is needed in order to absorb the twist.
- One can twist only in spatial directions to produce 1c; now fermions can be taken in flavor multiplets of 2. In the space directions the torus length in lattice units is odd and $N = 2\times$(a prime). In terms of the 0c larger $b$ phase boundary location, twisting has a similar effect in three dimensions.

As an example, we show a graph comparing Wilson loop spectra on a twisted $3^4$ lattice with $N = 28$ to the same kind of loops on an untwisted $9^4$ lattice at $N = 23$. In both cases $|b| = 0.36$, a value that cannot be attained on a substantially smaller untwisted lattice. The plot shows no marked difference, while the $3^4$ simulation is about 45 times cheaper. Also in three dimensions one obtains massive computer time reductions. This data is preliminary and the large saving in computer time could be accompanied by hidden problems, which could reduce the gain considerably. Nevertheless, we expect to make progress because there is a solid reason why twisting ought to help.

## 11. A few speculations

- In 0c twists may be adapted to simulations of the Veneziano limit ($N \to \infty$, $N_f \to \infty$, $\rho = N_f/N$=fixed) perhaps even at non-zero chemical potential, $\mu \neq 0$. Technically, there would be something to gain from the ability to turn on the imaginary part of the fermionic action gradually, by increasing $\rho$ from zero through a regime of weak non-hermiticity.

- If one does not reduce the time direction, reduction works for real time. More specifically, if we apply the Schwinger-Keldish formalism to states that are spatially homogeneous, large $N$ reduction ought to apply, at least to the point in real time that the an undesired $Z(N)$ symmetry breaking occurs dynamically. This indicates that one could employ reduction techniques to study nonequilibrium phenomena, a subject gaining importance in our subfield as new experimental results emerge from RHIC.

- The phases [2-4]c need to be studied: 3c likely is Bjorken's femptouniverse at infinite $N$ and 4c is Bjorken's femptouniverse at high temperature.

- There might be large $N$ phase transitions in 4D Wilson loops and in the 2D nonlinear chiral model: nontrivial eigenvalue dynamics could survive in the continuum limit. These transitions are not bulk large $N$ transitions as those separating the different Xc phases, but, rather some nonanalyticities occurring in particular operators as a function of their size. If these transitions do occur in the continuum limit it is a subtle issue how they can coexist with momentum analyticity. It is quite difficult to imagine how these transitions could be avoided altogether, and if they disappear in the continuum limit it might be important to understand their role on the lattice.

  In terms of Wilson loop operators one would like to claim that the renormalized continuum Wilson loop operators have the property that a backtracking portion of the loop $C$ drops out, as would be the case if the parallel transporter matrix were unitary. It that matrix is unitary, the information contained in its characters in all representations determines a spectrum on the unit circle. Once the spectrum is a meaningful continuum object, is is very hard to imagine anything else but a highly concentrated spectrum near eigenvalue unity for small loops and an almost uniform distribution round the entire circle for a very large loop, accounting for confinement. At infinite $N$ the spectrum ought to become continuous and for small loops we expect that the spectrum will have a gap centered at eigenvalue -1. For any finite $N$, the density of eigenvalues in the gap region would not vanish, but as $N$ is taken to infinity, experience teaches us that the spectrum goes to zero exponentially in $N$ in that region. We conclude that Wilson loops undergo a non-smooth transition in eigenvalue space as they are are dilated from small size. We have seen some examples of such transitions on the lattice, but a systematic study of the scaling behavior of the transition points has to wait for the future.

  In two dimensions the nonlinear chiral model has many analogs of four dimensional gauge theory. One can think of an analogue of the the above transitions too. One also has exact information about this two dimensional model, but that information does not tell us whether such transitions take place or not. We have started a numerical investigation in collaboration with L. Del Debbio and E. Vicari to study this issue.

## 12. Acknowledgments

This research was partially supported by the DOE under grant number DE-FG02-01ER41165 at Rutgers University and by the NSF under grant number PHY-030065 at FIU. Herbert Neuberger is grateful to the conference organizers for the invitation.